\documentclass[epj]{svjour}
\usepackage{graphics}
\usepackage{amssymb}
\usepackage{amsmath}
\begin{document}
\title{Near-threshold production of $\omega$ mesons in the
$pn\to d\omega$ reaction}

\author{
S.~Barsov\inst{1} \and
I.~Lehmann\inst{2} \and
R.~Schleichert\inst{2}\thanks{E-mail: r.schleichert@fz-juelich.de}\and
C.~Wilkin\inst{3} \and
M.~B\"uscher\inst{2} \and
S.~Dymov\inst{2,4} \and
Ye.~Golubeva\inst{5} \and
M.~Hartmann\inst{2} \and
V.~Hejny\inst{2} \and
A.~Kacharava\inst{6} \and
I.~Keshelashvili\inst{2,6} \and
A.~Khoukaz\inst{7}\and
V.~Komarov\inst{4}\and
L.~Kondratyuk\inst{8}\and
N.~Lang\inst{7}\and
G.~Macharashvili\inst{4,6}\and
T.~Mersmann\inst{7}\and
S.~Merzliakov\inst{4}\and
A.~Mussgiller\inst{2}\and
M.~Nioradze\inst{6}\and
A.~Petrus\inst{4}\and
H.~Str\"oher\inst{2}\and
Y.~Uzikov\inst{4}\and
B.~Zalikhanov\inst{4}}

\institute{
High Energy Physics Department, Petersburg Nuclear
Physics Institute, 188350 Gatchina, Russia \and
Institut f\"ur Kernphysik, Forschungszentrum J\"ulich,
D-52425 J\"ulich, Germany\and
Physics \& Astronomy Department, UCL, Gower Street,
London WC1E 6BT, UK\and
Laboratory of Nuclear Problems, Joint Institute for Nuclear
Research, Dubna, 141980 Dubna, Russia\and
Institute for Nuclear Research, Russian Academy of Sciences,
117312 Moscow, Russia\and
High Energy Physics Institute, Tbilisi State University,
University Street 9, 380086 Tbilisi, Georgia\and
Institut f\"ur Kernphysik, Universit\"at M\"unster,
W.-Klemm-Street 9, 48149 M\"unster, Germany\and
Institute for Theoretical and Experimental Physics,
Cheremushkinskaya 25, 117259 Moscow, Russia}

\date{Received: \today / Revised version: date}

\abstract{
The first measurement of the $pn\to d\omega$ total cross section
has been achieved at mean excess energies of $Q\approx 28$ and
$57\,$MeV by using a deuterium cluster-jet target. The momentum of
the fast deuteron was measured in the ANKE spectrometer at
COSY-J\"ulich and that of the slow ``spectator'' proton
$(p_\mathrm{sp})$ from the $pd\to p_\mathrm{sp}d\omega$
reaction in a silicon telescope placed close to the target. The
cross sections lie above those measured for $pp\to pp\omega$ but
seem to be below theoretical predictions.
 }
\maketitle
\section{Introduction}

The last few years have seen several measurements of $\eta$
production in nucleon-nucleon collisions~\cite{ppeta} but
relatively few of $\omega$ production~\cite{Hibou,TOF}. The
$S$-wave amplitude in the $\eta$ case is strong and the total
$pp\to pp\eta$ cross section largely follows phase space modified
by the $pp$ final state interaction up to an excess energy $Q =
\sqrt{s}-\sum_{f}m_f\approx 60\,$MeV, though there is some
evidence for an $\eta pp$ final state enhancement at very low
$Q\,$\cite{FJW}. Here $\sqrt{s}$ is the total centre-of-mass (cm)
energy and $m_f$ are the masses of the particles in the final
state. Quasi-free $\eta$ production in proton-neutron collisions
has been measured by detecting the photons from $\eta$ decay and
it is found that for $Q< 100\,$MeV the cross section ratio
$R=\sigma_\mathrm{tot}(pn\to pn\eta)/\sigma_\mathrm{tot}(pp\to
pp\eta)\approx 6.5$~\cite{Cal1}.  Now the $d\,\eta$ final state is
pure isospin $I=0$, whereas the $pp\eta$ is a mixture of $I=0$ and
$I=1$. Up to $Q\approx 60\,$MeV the cross section for $pn\to
d\eta$ is larger than that for $pn\to pn\eta$~\cite{Cal2}, and
this can be understood quantitatively in terms of phase space in a
largely model-independent way~\cite{FJW}. In all meson production
reactions it is important to have data on the different possible
isospin combinations in order to constrain theoretical models. It
is therefore interesting to see whether a similar isospin
dependence is found for the $\omega$, the next heavier isoscalar
meson.

Unlike the $\eta$ case, the $\omega$-meson has a significant width
(8.4$\,$MeV/c$^2$) and so $Q$ is here defined with respect to the
central mass value of 782.6$\,$MeV/c$^2$~\cite{PDG}. The $pp\to
pp\omega$ total cross section has been measured at five energies
in the range $4 \leq Q \leq 30\,$MeV at the SATURNE SPESIII
spectrometer~\cite{Hibou} and at $Q=92\,$MeV at
COSY-TOF~\cite{TOF} where, in both cases, the $\omega$ was
identified through the missing mass technique. The energy
dependence deduced is rather similar to that of the $\eta$, except
that the phase space and $pp$ final state interaction have to be
smeared over the finite $\omega$ width, a feature which becomes
important close to the nominal threshold~\cite{Hibou}.

Attempts to measure the $np\to d\omega$ reaction using a neutron
beam are complicated by the intrinsic momentum spread, which is
typically 7\% FWHM even for a stripped deuteron
beam~\cite{Sawada}. The alternative is to use a deuterium target
and effectively measure the momentum of the struck neutron. This
is made possible by detecting the very low momentum recoil
protons, $\lesssim 200\,$MeV/c, in the $pd\to p_\mathrm{sp}
d\omega$ reaction in a silicon telescope placed close to the
target. Such an approach is feasible at internal experiments at
storage rings such as CELSIUS or COSY because of the thin
windowless targets that can be used there. Under these conditions
the recoil proton can be largely treated as a ``spectator'' that
only enters the reaction through its modification of the
kinematics. The measurement of the fast deuteron in coincidence
would then allow us to identify the $\omega$ by the missing mass
method. By varying the angle and momentum of the spectator proton
it is possible to change the value of $Q$ while keeping the beam
momentum fixed. The principle of this method has been proved at
CELSIUS for the $pn\to d\pi^0$ reaction, where $Q$ could be
determined to 2$\,$MeV~\cite{Tord}.

\section{Experimental Set-up}

\begin{figure}[ht]
\begin{center}
\resizebox{0.4\textwidth}{!}{%
  \includegraphics{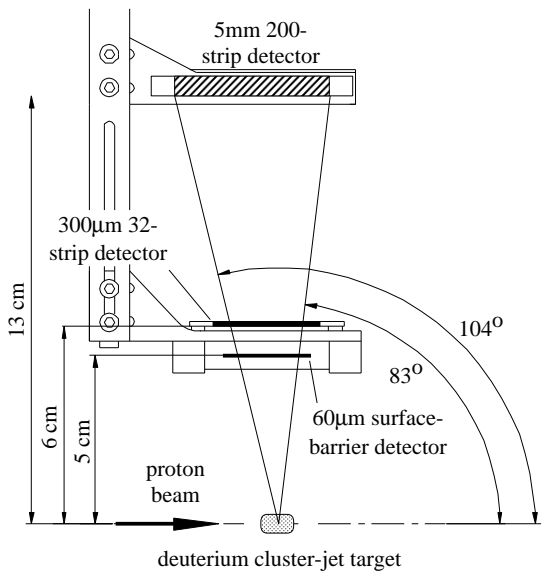}}
\caption{Top view of the telescope placed close to the
deuterium cluster-jet target inside the vacuum chamber of ANKE.
Protons and deuterons emerging from the beam-target overlap
region (shaded area) are detected in the subsequent arrangement
of three silicon detectors (see text).}
\label{SP_setup}
\end{center}
\end{figure}

Our experiment was performed using a deuterium cluster-jet
target~\cite{Mue} at the ANKE spectrometer~\cite{ANKE} situated inside
the COoler SYnchrotron COSY-J\"ulich, with the fast deuteron being
measured in the ANKE Forward Detector and the spectator proton in
solid state counters. The silicon telescope used for this purpose is
described in detail in Ref.~\cite{NIM_SP} and only the principal
features will be mentioned here. The three silicon layers indicated in
Fig.~\ref{SP_setup}, of respectively 60$\,\mu$m, 300$\,\mu$m, and
5$\,$mm, covered polar angles $83^{\circ}<\theta_{\mathrm{sp}}<
104^{\circ}$ and $\pm 7^{\circ}$ in azimuth. Protons with kinetic
energies $T_\mathrm{sp}$ in the range $2-6\,$MeV traversed the first
layer but were stopped in the second, while those in the range
$6-30\,$MeV were stopped rather in the final thick layer.  Energy
resolution of the order of $\sigma =150\,$keV was obtained. The
second and third layers were composed of strips arranged perpendicular
to the beam such that for $T_\mathrm{sp} > 8\,$MeV a resolution of
$\sigma(\theta_{\mathrm{sp}})\le 3^{\circ}$ could be achieved. For the
lower energy protons, neglecting the small non-target background, the
finite target size led to $\sigma (\theta_{\mathrm{sp}})\le
5^{\circ}$.

\begin{figure}[ht]
\begin{center}
\resizebox{0.45\textwidth}{!}{%
  \includegraphics{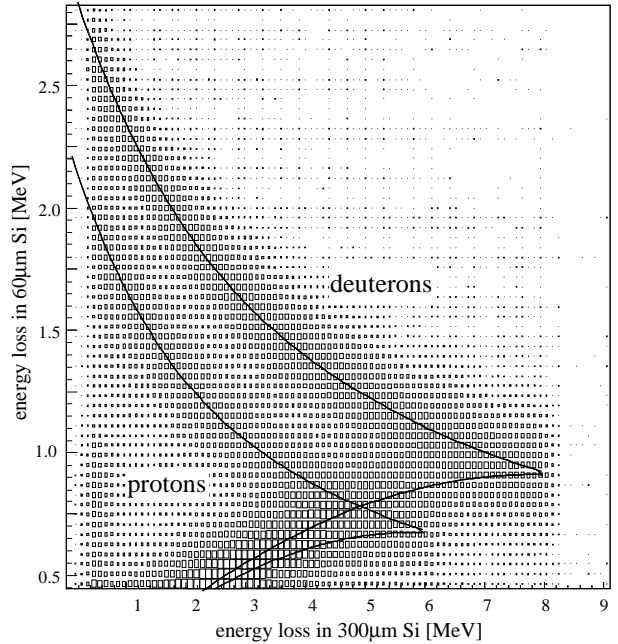}}
\caption{Particle identification {\em via} their energy deposit in
    the first two detectors of the telescope. Boxes corresponding
    to experimental data are compared to curves predicted for
        protons and deuterons.}
\label{SP_id}
\end{center}
\end{figure}

There was no difficulty in separating slow deuterons from protons
\textit{via} the $E-\Delta E$ method in two ranges: $2.6 <
T_\mathrm{sp} < 4.4\,$MeV ($70 < p_\mathrm{sp} < 91\,$MeV/c) and
also $8 < T_\mathrm{sp} < 22\,$MeV ($123 < p_\mathrm{sp} <
204\,$MeV/c). This is illustrated for the lower range in
Fig.~\ref{SP_id}. It is seen here that, by choosing the 4.4~MeV
upper limit, one avoids the possibility  of misidentifying
deuterons traversing the first two layers but missing the third.

The ability to identify a deuteron in the telescope in coincidence
with a proton in the forward detector also allows us to obtain
simultaneously the luminosity by measuring proton-deuteron elastic
scattering through a determination of the deuteron kinetic energy.
For this purpose we have calculated the elastic proton-deuteron
cross section at our energies within the Glauber
model~\cite{Glauber}. Such an estimate agrees with the available
experimental $pd\to pd$ data at 2.78$\,$GeV/c to within the quoted
error of about 10\%~\cite{pd_data}. The 1\% uncertainty in the energy
of the recoil deuteron, and hence in the momentum transfer,
induces only a 3\% error through the angular variation of the
normalising reaction. Due to uncertainties in the geometrical
constraints in the target chamber, the acceptance correction
introduces a 15\% systematic error in the absolute cross section
normalisation. The overall systematic luminosity error used to
determine absolute cross sections was thus taken to be 20\%. It
should be noted, however, that the error in the relative
normalisation between different beam energies is at most 5\%.

In order to distinguish deuterons with momenta around 2$\,$GeV/c,
arising from the $pd\to p_\mathrm{sp}d\omega$ reaction,
from a proton background that is two orders of magnitude higher,
inclined \v{C}erenkov counters were installed behind the
multi-wire proportional chambers and scintillator hodoscope of the
forward detection system of ANKE~\cite{ANKE,FD}. To understand the
detection principle, consider the detector response for a proton
and deuteron with the same momentum. The opening angle of the
\v{C}erenkov light cone for the faster proton is larger. Thus part
of the light can reach the photomultiplier after being totally
reflected in the counter, whereas all the light produced by the
deuteron leaves the counter. A momentum-dependent threshold was
applied so as not to change the differential distributions.

\begin{figure}[ht]
\begin{center}
\resizebox{0.45\textwidth}{!}{%
  \includegraphics{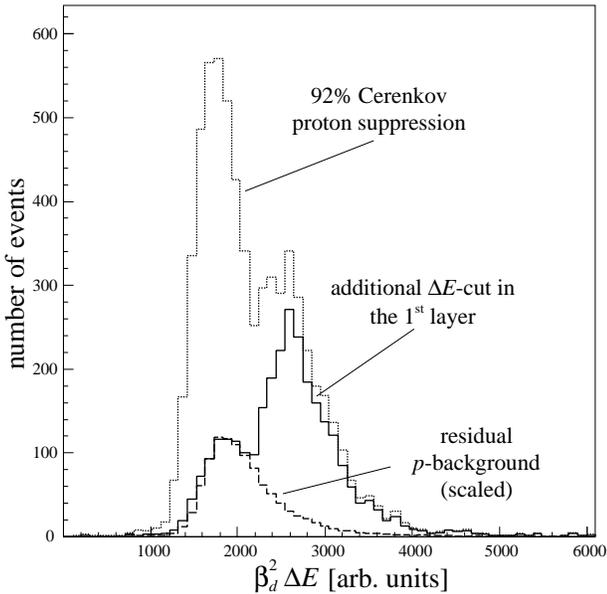}}
\caption{Energy loss in the second layer of the forward
scintillator hodoscope scaled with $\beta^{2}_{d}$, where
$\beta_{d}$ is the deuteron velocity. The dotted histogram shows
the result obtained after imposing a 92\% proton suppression with
the help of the \v{C}erenkov counters. The solid lines result
from adding a further $\Delta E$ cut using information from the
first layer. The residual proton background is shown by the dashed
line.}
\label{dE_proj}
\end{center}
\end{figure}

The hodoscope, consisting of two layers of scintillation counters,
provides an additional criterion for the deuteron identification
using the energy loss in both layers. By simultaneously varying
the $\Delta E$-cut and \v{C}erenkov efficiency level, an optimal
combination was found which leads to only a 20\% loss of deuterons
while giving a 92\% suppression of protons due to the \v{C}erenkov
counters alone. Projecting the energy loss in the second layer
along the predicted energy loss of deuterons ($\propto
\beta^{-2}_d$), one obtains the dotted histogram shown in
Fig.~\ref{dE_proj}. A further cut on the analogous distribution in
the first layer reveals a clear deuteron peak (solid line).
Moreover, the shape of the remaining proton background can be
determined using the energy loss distribution of suppressed
particles which, after scaling, is drawn as the dashed line. This
shows that the proton background is on the 10\% level.

\section{Data Analysis}

Having identified a spectator proton in the telescope and a
deuteron in the forward array and furthermore measured their
momenta and directions, one can evaluate the missing mass $m_X$ in
the reaction. To clarify the effects of the kinematics, it is
sufficient to treat the spectator as being non-relativistic. To
order $p_{\mathrm{sp}}^2$ we have then
\begin{equation}
m_X^{\,2} \approx \tilde{m}_X^{\,2} + 2(\mathbf{p}_d-\mathbf{p})\cdot
\mathbf{p}_{\mathrm{sp}} - 2\left(\frac{E+m_d-E_d}{m_p}\right)
\,p_{\mathrm{sp}}^2\:,
\end{equation}
where $\tilde{m}_X$ is the value obtained at
$p_{\mathrm{sp}}=0$. Here $p$ and $E$ are the laboratory
momentum and total energy of the incident proton, $p_d$ and $E_d$
those of the produced deuteron, and $m_d$  and $m_p$ the masses of
the deuteron and proton respectively. The square of the $pn$ cm
energy, $s$, can be evaluated purely using measurements in the
spectator counter and from this $Q$ can be derived:
\begin{equation}
s=(m_d+m_{\omega}+Q)^2
\approx\tilde{s} + 2pp_{\mathrm{sp}}\cos\theta_{\mathrm{sp}}
-\left(\frac{E+m_d}{m_p}\right)\,p_{\mathrm{sp}}^2\:,
\end{equation}
where $\tilde{s}$ is the value for a stationary neutron. Because
the telescope is placed around $\theta_{\mathrm{sp}}\approx
90^{\circ}$, $\partial s/\partial\theta_{\mathrm{sp}}$ is then
maximal and so the value of $Q$ depends sensitively upon the
determination of the polar angle of the spectator with respect to
the beam direction.

Since in our set-up the fast deuteron is measured near the forward
direction, the same sort of sensitivity is also found for $m_X$ when
using Eq.~(1). Now for each beam momentum the beam direction could not
be established to much better than $0.1^{\circ}$, and this may induce
a systematic shift of a few MeV/c$^2$ in the value of $m_X$.
On the other hand, in view of the
$\omega$ width, the uncertainty in the beam momentum ($< 1\,$MeV/c) is
unimportant for both $Q$ and $m_X$ at this level of accuracy. The
struck neutron is slightly off its mass shell but the off-shellness is
controlled by the spectator momentum and rests small throughout our
experiment.

\begin{figure}[ht]
\begin{center}
\resizebox{0.45\textwidth}{!}{%
  \includegraphics{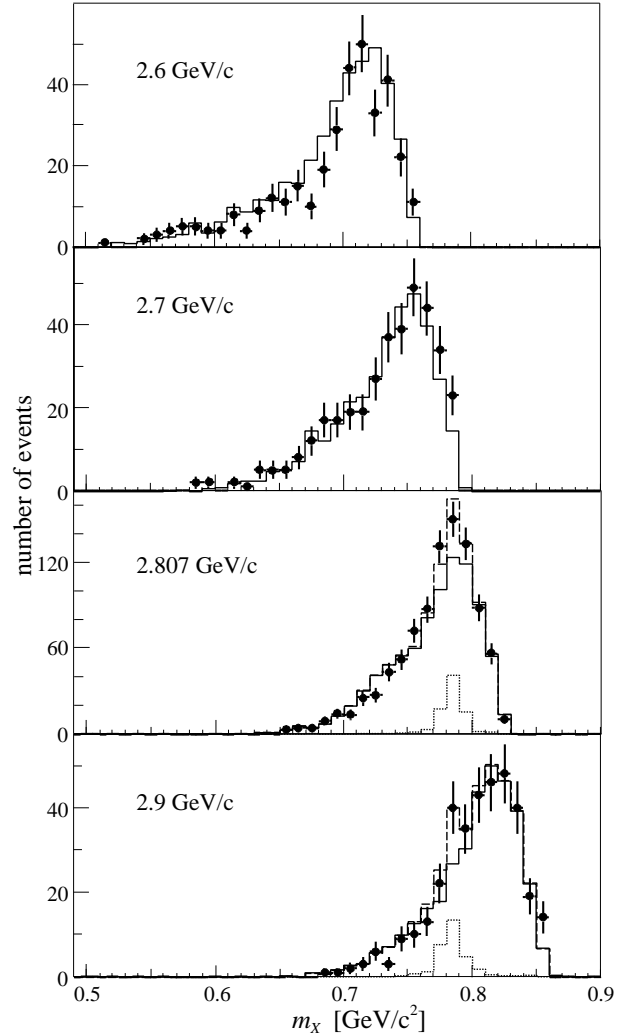}}
 \caption{Missing mass $m_X$ in the $pd\to p_\mathrm{sp}dX$
          reaction. Two of the four beam momenta 2.6, 2.7, 2.807,
          and 2.9$\,$GeV/c are largely below and two above the nominal
          threshold for $\omega$-production, with excess energies summed
          over the ranges $-58<Q<-22$, $-23<Q<13$, $8<Q<44$, and
          $42<Q<78$~MeV respectively. The solid line is a multipion
          fit to the data at 2.7$\,$GeV/c and 2.9$\,$GeV/c, as described
          in the text. There is evidence for an $\omega$ signal (simulated
          as the dashed line) at 2.9$\,$GeV/c, though the result at
          2.807$\,$GeV/c depends much more sensitively upon the background
          simulation.
    }
 \label{f:mx}
\end{center}
\end{figure}

In Fig.~\ref{f:mx} we show our results from the first two silicon
layers ($70 < p_\mathrm{sp} < 91\,$MeV/c), where the spectator
hypothesis should be very good. The angular information is
important for the missing mass determination but, in view of the
limited statistics, we had to sum over rather wide bins in excess
energy. Experience with $\omega$ production in proton-proton
collisions shows that there is considerable multi-pion production
under the $\omega$ peak~\cite{Hibou}. Without measuring the
products of the $\omega$ decay, this can only be reliably
estimated by comparing data above and below the $\omega$
threshold. Two of the four momenta correspond to largely
below-threshold measurements and two above, at mean values of
$Q$ equal to about 28 and 57$\,$MeV.

There is an indication of a weak $\omega$ signal at the highest
energy and, in order to evaluate its significance, we have to
master the large multipion background over our range of energies.
Two different approaches have been undertaken to overcome this
problem. In the first, pion production is modelled within a
phase-space Monte Carlo description. The second method is
identical to that used in the analysis of the $pp\to pp\,\omega$
experiment~\cite{Hibou}, where the data below $\omega$ threshold
were taken to be representative of the background above, being
merely shifted kinematically due to the changed beam energy such that
the upper edges of phase space match. This matching of the ends of
phase space can also be used to check the set-up of the system at each
momentum.  The only significant discrepancy was found
at 2.807$\,$GeV/c where, in order to account for a slight displacement
observed in the data, 3$\,$MeV/c$^2$ has been subtracted from all
$m_X$ values at this beam momentum. As
will be shown in the next section, the two different analysis
methodologies give consistent results within the error bars.

Most of the background can be described by phase space convoluted
with the ANKE acceptance, which provides a severe cut at low
$m_X$. It
should be noted that the available $np\to d\pi^+\pi^-$ data in our
energy range show the deuteron distribution to be fairly isotropic
in the cm system~\cite{Abd}.
In the absence of neutron data, we parameterised the total
cross section $\sigma(s)$ for the production of $N$ pions in
proton-proton collisions by
\begin{equation}
\sigma(s) =
A\,\left(1-\frac{s_0}{s}\right)^{p_1}\,
\left(\frac{s_0}{s}\right)^{p_2}\:, \label{ENERGY_DEPENDENCE}
\end{equation}
where $s_0$ is the threshold for $N\pi$ production. The exponent
$p_1$ is fixed by phase space, but $A$ and $p_2$ are free
parameters adjusted to reproduce the $pp\to d(N\pi)$ data for 2, 3
and 4 pion production~\cite{pi_data}. The assumption that each of
the three contributions follows a $(N+1)$-body phase space,
undistorted by $\Delta$ or $\rho$ resonances, gives a description
of the $m_X$ distributions for different beam energies. To model
the $pn\to d(N\pi)$ background, the energy dependence from the
$pp$ case has been used to fix the $p_i$, with the $A$ being
adjusted to reproduce simultaneously our experimental distribution
at 2.7$\,$GeV/c and the phase-space maximum at 2.9$\,$GeV/c.
The relative normalisation between these two momenta was determined
from the $pd$ elastic scattering data.
The adjusted A values, together with the relative normalisation
established from the luminosity measurement, were used to describe
the multi-pion background at 2.6$\,$GeV/c and at 2.8$\,$GeV/c,
as shown in Fig.~\ref{f:mx}.

Our method gives a plausible description of the background under
the $\omega$ peak at $Q\approx 57\,$MeV but any $\omega$ signal at
$Q\approx 28\,$MeV lies close to the maximum of the phase-space
acceptance and the evaluation of its strength depends much more
critically upon the background assumptions. Nevertheless, within
the parametrisation of Eq.~(3), it is impossible to describe the
phase-space maxima simultaneously at the four energies in
Fig.~\ref{f:mx} without invoking some $\omega$ signal at $Q\approx
28\,$MeV.

To describe the $\omega$ contribution to the missing mass spectra, we
take the $pn\to d\omega$ matrix element to be constant over the
$Q$-bin so that the cross section follows phase space. This, combined
with the decrease of acceptance at large $Q$, means that the mean
value of $Q$ is not quite at the centre of the bin. Other plausible
assumptions, such as a constant cross section, would lead to
negligible changes in the evaluation of the cross section and mean
value of $Q$.  In the simulation of the $pd\to p_\mathrm{sp}d\omega$
reaction, the cross section is smeared over the Fermi motion in the
deuteron using the PLUTO event generator~\cite{PLUTO}. This employs
the Hamada-Johnston wave function~\cite{DWF} though, at these small
values of spectator momenta, other more realistic wave functions give
indistinguishable results. The same event generator is used also for
the multipion background.

\begin{figure}[ht]
\begin{center}
\resizebox{0.45\textwidth}{!}{%
  \includegraphics{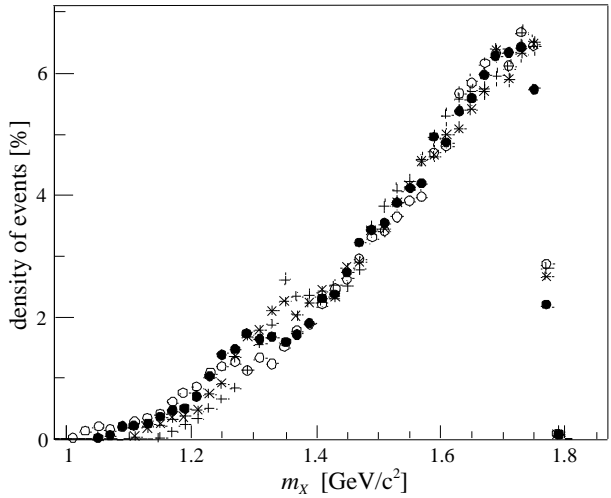}}
 \caption{Missing-mass spectra of the $pd\to p_\mathrm{sp}p\,X$
 reaction at 2.6 (crosses), 2.7 (stars), 2.8 (closed circles), and
 2.9~GeV/c (open circles) kinematically shifted using the SPESIII
 procedure~\cite{Hibou}. The data are all normalised to the same
 total of 100\%.}
 \label{p-spectra}
\end{center}
\end{figure}

Turning now to our second approach, the authors of
ref.~\cite{Hibou} noticed that, apart from the $\omega$ signal,
the shape of the $pp\to pp\,X$ missing mass spectrum varied little
with beam energy provided that one looked at the distribution with
respect to the maximum missing mass. More quantitatively, if
$\beta$ and $\beta_n$ are c.m.\ velocities at energies $T$ and
$T_n$ respectively, the measured momenta and angles of the protons
were first transformed, event-by-event, from the laboratory to the
c.m.\ system with the velocity $-\beta$ and then transformed back
to the laboratory with the velocity $+\beta_n$. To see to what
extent this approach is valid for the ANKE spectrometer, which has
a much smaller overall acceptance than that of
SPESIII~\cite{Hibou}, we have reconstructed the missing mass for
the copious proton production $pd\to p_\mathrm{sp}p\,X$. The data
at the four different beam momenta, kinematically shifted to
2.9$\,$GeV/c and normalised to the same total number of events,
are shown in Fig.~\ref{p-spectra}.

It is clear from the figure that for $m_X>1.4$~GeV/c$^2$ the
shifted data are in mutual agreement at all beam momenta. For
lower missing masses one sees the effect of the production of the
$\Delta(1232)$ isobar, whose position in the \textit{shifted} mass
scale depends, of course, upon the beam momentum. The figure also
nicely illustrates the influence of the ANKE acceptance cut, which
strongly favours events close to the maximum missing mass.

\begin{figure}[ht]
\begin{center}
\resizebox{0.45\textwidth}{!}{%
  \includegraphics{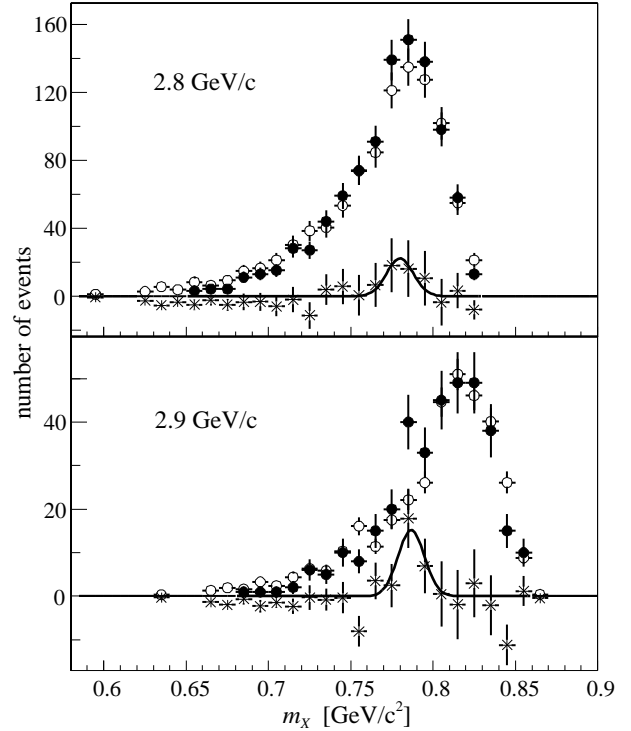}}
 \caption{Missing-mass spectra of the $pd\to p_\mathrm{sp}d\,X$
 reaction. The closed circles are the experimental data at 2.8
 and 2.9~GeV/c whereas the open circles represent the data
 at the other momenta shifted using the SPESIII
 procedure~\cite{Hibou}. The differences between the data sets
 (stars) are fitted to the expected $\omega$-peak shape to yield
 the measured production cross section.}
 \label{d-spectra}
\end{center}
\end{figure}

When the identical analysis procedure is applied to the $pd\to
p_\mathrm{sp}d\,X$ data, the backgrounds away from the $\omega$
peak at the different momenta are again found to be completely
consistent. An average background could therefore constructed and
this is shown for the two above-threshold momenta in
Fig.~\ref{d-spectra}. The differences between the experimental
data and constructed background shows evidence for structure in
the $\omega$ region and these have been fitted to $\omega$ peaks
whose widths were fixed by the Monte Carlo simulation. The
$\omega$ masses obtained from the fits at the two momenta,
$780\pm8$ and $787\pm 4$~MeV/c$^2$, do not differ significantly
from the expected value.

\section{Results}

By comparing the residual signal in Fig.~\ref{f:mx} with a
simulation of $\omega$ production over this range of spectator
energies and angles, we would conclude from simulated background
model that $\sigma_\mathrm{tot}(pn\to d\,\omega)=(2.9\pm
0.8)\,\mu$b at $Q=(28^{+16}_{-20})\,$MeV and $(8.5\pm 2.8)\,\mu$b
at $Q=(57^{+21}_{-15})\,$MeV, where the uncertainty in $Q$
reflects the total width of the bin and only the statistical error
in the cross section is quoted. The corresponding numbers obtained
using the SPESIII background technique, $(2.2\pm 1.4)\,\mu$b and
$(9.4\pm 3.3)\,\mu$b respectively, are consistent with the first
method, though the statistical errors are larger because we had to
subtract a background with limited statistics. This contrasts with
our first approach where we imposed the condition that the
background should be smooth. Averaging the two sets of results, we
obtain $\sigma_\mathrm{tot}=(2.6\pm 1.6\pm2.3)\,\mu$b and $(9.0\pm
3.2^{+3.6}_{-2.5})\,\mu$b at the two excess energies. The second,
systematic, error bar includes some contribution arising from the
ambiguity of the background discussed above but others, such as
the uncertainty in the luminosity, are common to both the signal
and background.

In view of the limited statistics it might be helpful to quote
upper limits resulting from the fits to the count differences
shown in Fig.~\ref{d-spectra}. At the 90\% confidence level the
cross sections at 2.8 and 2.9~GeV/c are below $(7.5\pm 5)~\mu$b and
$(17\pm 6)~\mu$b respectively, where the second figure is the
rescaled systematic uncertainty.

\begin{figure}[ht]
\begin{center}
\resizebox{0.45\textwidth}{!}{
  \includegraphics{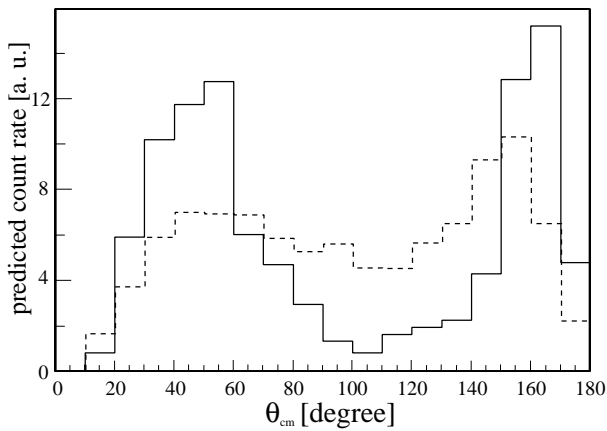}}

 \caption{Predicted angular acceptance for $pd\to
 p_\mathrm{sp}d\,\omega$ events at 2.9$\,$GeV/c (solid line) as a
 function of the deuteron c.m.\ angle, assuming an isotropic
 production process. At 2.8$\,$GeV/c (dashed line) the distribution
 becomes more uniform.}

 \label{theta_dw}
\end{center}
\end{figure}

One source of systematic uncertainty comes from the restricted
angular acceptance of ANKE~\cite{ANKE}, a problem that becomes
more serious with increasing $Q$. The simulation of the
acceptance, illustrated in Fig.~\ref{theta_dw} for an isotropic
production distribution, shows that, while the distribution is
fairly flat at 2.8$\,$GeV/c, few events would be accepted close to
$90^{\circ}$ at 2.9$\,$GeV/c. Although at our energies we might
expect $S$-wave production to dominate, when this acceptance is
weighted with the possible pure $P$-wave angular variations of
$\cos^2\theta$ or $\sin^2\theta$, the resulting overall acceptance
estimate at 2.9$\,$GeV/c is changed by factors of of 1.7 and 0.65
respectively. These are, however, extreme scenarios and a
systematic error of half of the difference between these values is
a generous estimate of this uncertainty. In more refined
experiments, where the statistics will allow us to determine the
angular distribution, this limitation will be avoided.

\begin{figure}[ht]
\begin{center}
\resizebox{0.45\textwidth}{!}{%
  \includegraphics{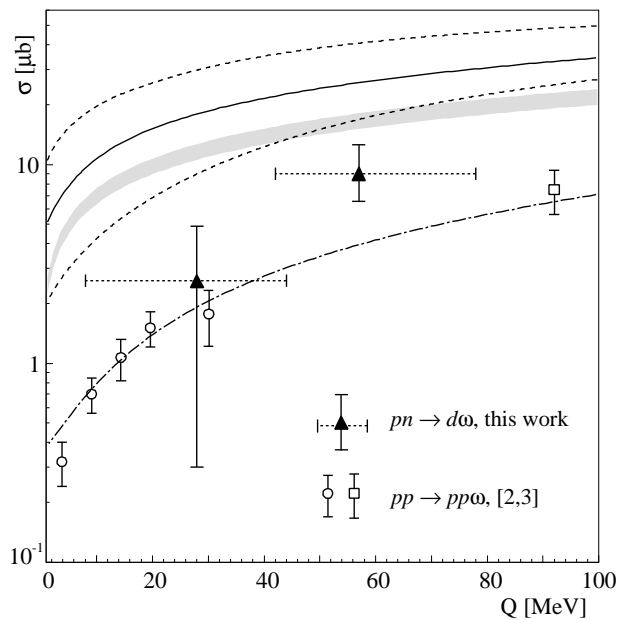}}

\caption{Total cross sections for $\omega$-production. The $pp\to
pp\omega$ data are taken from SATURNE~\protect\cite{Hibou} (open
circles) and COSY-TOF~\protect\cite{TOF} (open square), whereas our
two $pn\to d\omega$ points are given by the closed triangles. Only the
systematic errors are shown as the statistical errors 1.6 and
3.2$\,\mu$b at the two energies are smaller or comparable in size. The
horizontal bars indicate the width of the $Q$ ranges. The dot-dashed
curve is the semi-phenomenological fit given in
Ref.~\protect\cite{Hibou} to the $pp\to pp\omega$ results taking the
$\omega$ width into account. If the ratio for $d\omega$ to $pp\omega$
were similar to that for $\eta$ production~\cite{Cal2}, one would then
obtain the solid curve, which predicts a $pn\to d\omega$ cross section
of over 25$\,\mu$b at 57$\,$MeV. The predictions of the J\"ulich group
depend upon the relative contributions of exchange and production
current terms and lie between the two dashed
curves~\protect\cite{JTG}. The only other published
estimate~\protect\cite{Grishina} is shown by the shaded area.}%

 \label{f:points}
\end{center}
\end{figure}

The reduction of flux due to the presence of a second nucleon in
the deuteron target (shadowing) has been estimated in the
$\eta$-production case to be about 5\% of the $NN$ cross
section~\cite{PINOT} and such a correction has been applied to our
data. These values are shown in Fig.~\ref{f:points} along with
those for the $pp\to pp\omega$ reaction.

Due to the momentum distribution in the deuteron, the statistics
for spectators in the higher range, $8 < T_\mathrm{sp} < 22\,$MeV,
are only about a third of those in the lower range. Nevertheless,
the corresponding missing mass spectra are consistent with those
shown for the lower spectator energies in Fig.~\ref{f:mx}, with
$\omega$ cross sections compatible with our results in
Fig.~\ref{f:points}.

We have checked our methodology by identifying events
corresponding to the $pd\to p_\mathrm{sp} d\pi^0$ reaction at
1.22$\,$GeV/c. Using the same procedures as for the $\omega$
analysis, we find $\sigma_\mathrm{tot}(pn\to d\pi^0) = (1.6\pm
0.3)\,$mb at $Q=(135\pm 9)\,$MeV, where the statistical error is
negligible. This value is to be compared to $1.53\,$mb deduced
from a compilation of the isospin-related $pp\to d\pi^+$
reaction~\cite{SAID}.

\section{Conclusions}

In any meson exchange model, the relative strength of $\omega$
production in $pp$ and $pn$ collisions depends sensitively upon
the quantum numbers of the exchanged particles. If only a single
isovector particle, such as the $\pi$ or $\rho$, were exchanged
then, neglecting the differences between the initial and final
$NN$ interactions, one would expect $\sigma_\mathrm{tot}(pn\to
pn\omega)/\sigma_\mathrm{tot}(pp\to pp\omega)=5$. This would
explain most of the 6.5 factor found in the $\eta$
case~\cite{Cal2}. Assuming that the ratio $d\omega$ to $pp\omega$
is as for $\eta$ production, the parametrisation of the available
$pp\to pp\omega$ data~\cite{Hibou,TOF} leads to the solid curve,
which lies about a factor of three above our data. Another
estimate is a little lower but similar in shape~\cite{Grishina}.
Both curves lie within the extremes of the predictions of the
J\"ulich theory group~\cite{JTG}, where the major uncertainty
arises from the relative strengths of production and exchange
current terms.

Taking our 90\% C.L.\ upper limit on the cross section, augmented
by the corresponding systematic uncertainty, would barely
bring the data into agreement with the solid line of
Fig.~\ref{f:points}. Even considering only these upper limits,
the model predictions appear higher than the data.
Any theoretical overestimation might be
explained if there were significant isoscalar exchange, perhaps
through the $\omega$ itself.

In summary, we have carried out the first measurement of the
$pn\to d\omega$ reaction by detecting the spectator proton from a
deuterium target in coincidence with a fast deuteron. Although the
data are of very limited statistical significance, they suggest that the cross
section lies below the published theoretical predictions.

In order to clarify the situation further, we are constructing
second generation silicon telescopes that will increase the
acceptance significantly. It would then be of interest to try to
extend this study to the $\phi$ region so that one could
investigate the OZI rule in the $I=0$ channel to see if the
deviations are similar to those in the $I=1$ channel.

The work reported here formed part of the PhD thesis of one of the
authors~[IL]. We are most grateful to the team of the IKP
semiconductor detector laboratory, D.~Proti\'c, T.~Krings and
G.~Fiori, who developed and supported the necessary material for
our spectator counters. We wish to thank D.~Prasuhn, J.~Stein,
B.~Lorentz and the COSY team for carrying out the beam development
required by the condition that our detectors are only 5$\,$cm from
the circulating proton beam. P.~Wieder, W.~Borgs and
S.~Mikirtichiants helped in preparing and performing the
experiment. The work has been financially supported by the DFG
(436 RUS 113/630/71), the Russian Academy of Science
(RFBR02-02-0425) and the FZ-J\"ulich (COSY-064).

\end{document}